# Coordination Technology for Active Support Networks: Context, Needfinding, and Design[*]

November 6, 2017

Stanley J. Rosenschein
BranchTime Technologies
Palo Alto, CA 94301 USA
*and*
Center for the Study of Language and Information
Stanford University
stan@branchtime.com

Todd Davies
Symbolic Systems Program
Stanford University
Stanford, CA 94305 USA
*and*
Center for the Study of Language and Information
Stanford University
davies@stanford.edu

**Abstract**

Coordination is a key problem for addressing goal-action gaps in many human endeavors. We define interpersonal coordination as a type of communicative action characterized by low interpersonal belief and goal conflict. Such situations are particularly well described as having collectively "intelligent", "common good" solutions, *viz.*, ones that almost everyone would agree constitute social improvements. Coordination is useful across the spectrum of interpersonal communication – from isolated individuals to organizational teams. Much attention has been paid to coordination in teams and organizations. In this paper we focus on the looser interpersonal structures we call active support networks (ASNs), and on technology that meets their needs. We describe two needfinding investigations focused on social support, which examined (a) four application areas for improving coordination in ASNs: (i) academic coaching, (ii) vocational training, (iii) early learning intervention, and (iv) volunteer coordination; and (b) existing technology relevant to ASNs. We find a thus-far unmet need for personal task management software that allows smooth integration with an individual's active support network. Based on identified needs, we then describe an open architecture for coordination that has been developed into working software. The design includes a set of capabilities we call "social prompting," as well as templates for accomplishing multi-task goals, and an engine that controls coordination in the network. The resulting tool is currently available and in continuing development. We explain its use in ASNs with an example. Follow-up studies are underway in which the technology is being applied in existing support networks.

**Keywords**

coordination, social prompting, communicative action, attitude-behavior gap, intention-action gap, active support networks

**Acknowledgements**

We have benefited from conversations with Jeff Shrager, Rich Levinson, and Marc Smith, as well as an email exchange with Doug Schuler and the comments of two anonymous reviewers.

---

[*] Scheduled to appear in *AI & Society,* Special Issue on Collective Intelligence for the Common Good, Volume 33, Number 1, January 2018

# 1. Introduction

There is a well-recognized gap between goals and behavior in human action, generally, but especially in the realm of prosocial behavior.[1] Many would affirm the importance of making a personal contribution to the wellbeing of others, especially those most in need, but would also admit to not acting on those attitudes as often as they would like. They might respond to a few specific appeals over the course of a year, e.g., to donate money to a good cause or sign a petition, but would be less likely to involve themselves in a sustained way that drew on their personal skills, time, and efforts. If the gap between goals and action could be closed or even significantly reduced, society would be well served and the lives of many people improved.

While much research on bridging the goal-action gap focuses on motivational factors (e.g. Godin, Conner, & Sheeran, 2005), technology solutions may be particularly well suited to addressing the more cognitive challenges of managing attention, task suspension and resumption, and coordination among cooperating individuals.[2] The approach described in this paper does not assume that the rise of digital technology will inevitably amplify human abilities to bring about the common good, nor that "human nature" will sabotage such outcomes. Instead, we will explore how empirically-driven tool development may bring about dramatic improvements in follow-through and coordination, which can then be invested in socially beneficial outcomes.

# 2. Coordination, Communicative Action, and CI4CG

Our understanding of "coordination" is that it is a crucial component of what has been referred to as "collective intelligence for the common good" or "CI4CG" (De Cindio *et al.*, 2014). We will define coordination within the broader category of "communicative action" -- an umbrella term that encompasses "collaboration," "deliberation," and other concepts associated with collective intelligence. The term "communicative action" was introduced by Jürgen Habermas, who developed an extensive theory around it (Habermas, 1984; Habermas, 1987). For the purposes of this paper, we take "communicative action" to mean a joint action in which participants communicate, and in which the action is the purpose to which the communication is directed.[3]

We can define different forms of communicative action in terms of the types and levels of conflict between participants that the forms are geared toward resolving. Table 1 places four forms of communicative action in the space of conflict over beliefs and preferences.

When both belief and preference conflict are low, participants agree on the state of the world and the consequences of alternative actions, and also on what they are trying to achieve. Their task in this part of the conflict space is *coordination*, and their main need is for clear, efficient communication. This is the classic case of *collaboration*. When preference conflict is low but belief conflict is high, participants are apt to engage in *argumentation*. A systematic approach to resolving the belief conflict in this situation is that of "adversarial collaboration" (Tetlock & Mitchell, 2009).

---

[1] This gap is not limited to prosocial behavior, but exists whenever short-term or personal needs/desires interfere with longer-term or broader goals. The gap can be between attitudes or values and behavioral intentions (see e.g. Vermeir & Verbeke, 2006), or between intentions and actual behavior (Sniehotta, Scholz, & Schwarzer, 2005; Kuo & Young, 2008).
[2] Cognitive challenges stem especially from the limitations of human attention and memory, which tend to be overwhelmed in the age of information. See Levitin (2014).
[3] Habermas adds: "communicative action designates a type of interaction that is coordinated through speech acts and does not coincide with them" (Habermas, 1984, p. 101). This distinguishes communicative action from speech acts, because in a speech act, the speech *is* the action (Searle, 1969).

Table 1 Four Forms of Communicative Action

|  |  | *Preference Conflict* | |
|---|---|---|---|
|  |  | *low* | *high* |
| **Belief Conflict** | *low* | coordination | negotiation |
|  | *high* | argumentation | deliberation |

Social actors may, alternatively, be in low belief conflict but high preference conflict -- a situation often best resolved through *negotiation*. In this case, participants agree about the consequences of future actions, but have different preferences among alternatives and must typically compromise. The support they need in this situation may involve technology that helps participants see possibilities such as Pareto-improving trades and the Best Alternative to a Negotiated Agreement (BATNA), with research showing that unaided bargains are often suboptimal (Bazerman & Neale, 1992).

The most challenging region in the space of communicative action is where *both* beliefs and preferences are highly in conflict between participants. This is the usual case in *deliberation*, when actors with competing perceptions and interests must reach a collective decision. Deliberation is challenging because participants often cannot see whether conflict arises from competing goals or divergent beliefs, and they can even have an incentive to misrepresent their true feelings on one or both dimensions. Technology support for deliberation is a relatively new but growing area of design, research, and practice (Davies & Gangadharan, 2009).

In this paper we focus just on the lowest conflict cell in Table 1, what we call "coordination". Coordination is a good starting place for looking at how technology can improve collective intelligence for the common good, because it is the simplest of the four cells (with applications to the others), and because the low level of interpersonal conflict that defines it seems most amenable to solutions that merit the term "intelligent". "Smart" and "intelligent" in the realm of social outcomes apply least controversially when goals (and to a lesser extent) beliefs do not differ fundamentally across individuals for a particular set of outcomes.[4] This seems closest to the idea of the "common good" as "a universal benefit, something that everybody - in theory -- would want" (quote from De Cindio *et al.*, 2014).

## 3. Goals and Actions in Coordination: Individuals, Networks, and Teams

The processes involved in decision-making and action have been studied at many levels by a range of specialists, from neuroscientists to social scientists to organizational theorists.[5] For our purposes, a simple model will suffice to illuminate the special problems involved in designing technology that amplifies human decision and action, and therefore coordination:

(1) Humans carry around, at any point in time, a collection of goals (manifested as commitments, intentions, tasks, etc.).
(2) These goals represent outcomes they plan to bring about, though they may not yet know how.
(3) Many of these goals are private, known only to the individual. Some may be known by others, and that knowledge may play into intricate behaviors of support or competition.
(4) From time to time, the individual engages with a goal, bringing it to mind, and perhaps taking an action in the external world (e.g. picking up a briefcase, placing a phone call) or perhaps making a decision about what, when, or how to accomplish a portion of the goal.
(5) When goals are supported by the decisions and actions of others, communication among the relevant individuals accompanies these individual decisions and actions.

---

[4] This idea echoes critiques by Greenfield (2013) and Sadowski (2016) of the concept of intelligence as applied to "smart cities": that many/most policies and practices put in place in a city will benefit some citizens and harm others, or will accord with the beliefs of some but not all, so that in such cases there is no agreed upon standard for measuring whether a given action is "smart" in the context of a city. At the level of deep philosophy, Arrow's impossibility theorem (Arrow, 1963) and related results in the theory of social choice undergird an argument that the aggregation of preferences (and therefore of goals) across individuals cannot be performed in a way that produces collective rationality, *viz.*, a single preference ordering that is well-ordered and consistent for any possible set of individual preferences, without resorting to dictatorial imposition that may be at odds with what most people want. For these reasons, we think it is best to view "intelligence" as a consensus concept, and to limit our use of "collective intelligence" (as opposed to intelligence exhibited individually by multiple people) to situations where the individuals' beliefs and preferences can at least eventually come into rough alignment with respect to a particular set of outcomes.

[5] For an overview of the classic work, see Baron (2007).

How does technology enter this picture? Because many episodes of recurrent engagement may be needed for all the requisite decisions and actions to be taken, some mechanism is needed to ensure reliable and timely attention to the "next step." In simple cases, our inborn recall suffices. In other cases, we occupy a cue-rich environment where the stimuli we encounter (including incoming communications) trigger or refocus attention on the "next step." In yet other situations, if the goal is sufficiently complex or the environment is insufficiently cue-rich, we may design "systems" that ensure timely re-engagement. This is often the case in organizations, project teams, and the paper- and computer-based systems that support them.

Many individuals have adopted a personal system (electronic or otherwise) to help them re-engage with processes and goals in their lives. Behavioral scientists have popularized the concept of nudges (Sunstein & Thaler 2008), checklists (Gawande 2010), behavior triggers (Fogg & Hreha 2010), and commitment devices (Rogers, Milkman, & Volpp, 2014) to address specific challenges in behavioral intervention and effectiveness. We wish to build on this work in addressing interpersonal coordination for the common good.

We can distinguish different levels and sub-levels of interpersonal engagement that requires coordination:

- *Individuals* engage in behavior that requires coordination between tasks, even if the individual is viewed as *isolated* from others, and that requires coordination with others when that individual is *connected* to them;
- *Networks* of individuals, who are connected to each other but who retain autonomy to decide on their own actions, may coordinate for *passive support* or *information-sharing*, e.g. in communities of practice (Wenger, 1998), or to provide more direct *active support* for each other; and
- *Teams*, whose members are consciously working together to achieve the same goal(s), and must typically coordinate with each other whether they come together on an *ad-hoc* basis or as members of an *organization*.

In this paper, we focus on networks, and specifically on networks of active (as opposed to passive, information-sharing) support, or "active support networks" ("ASNs"). We define these as loose (open or shifting membership) collections of individuals that may include both peers and those who are coaching or assisting them, to achieve particular goals or deal with ongoing problems. Active support networks are distinct from passive support networks, in which members share knowledge and skills but do not typically take an active role in each others' activities, on one hand, and on the other hand, from ad-hoc or organizational teams, which are bound together by a shared, common obligation and explicit membership in a team. Individuals in an ASN must often coordinate with each other to get things done, but their motivation for doing so must typically come from their own individual goals or from a desire to help others, rather than from a structure that will hold them accountable. In our application of this concept, we will include mixed networks of support in which some members have a professional obligation to help others or to get something done, but they must rely on and coordinate with others who are not under the same obligations and so may be viewed either as volunteers or as individually motivated.

## 4. Needfinding Investigations

In this section, we describe two sets of investigations aimed at discovering what technology is needed for coordination in active support networks of the kind described in section 3. We first report on what kinds of software is needed for four particular types of ASNs (application areas), with particular attention to mixed professional-volunteer ASNs. We then report on an investigation of existing software that is plausibly relevant to these types of ASNs, in order to determine whether existing tools can meet the needs identified in our sample of application areas.

### 4.1 Exploration of Application Areas

We began by focusing on the domain of mixed professional-community active support networks. These networks have the crucial characteristics of an ASN (see section 3) in that their members are not bound together by a formal structure of accountability, and yet they engage in activity that requires coordination. There are many professions where success requires active cooperation (and even collaboration) between social-service professionals and the individuals or communities they serve. Examples include, broadly, practitioners and case managers in community services, educational professionals, and healthcare professionals. Many of the situations dealt with by these professionals require specific behaviors on the parts of individuals who are manifestly not part of the institutions the professionals represent. Typically, practitioners attempt to influence behavior through ordinary communication channels: in-person office visits, parent-teacher meetings, home visits by social workers, and so on. This personalized guidance may be extended to occasional phone calls and emails, and supplemented by pointers to relevant educational materials, in books, websites, and so on.

We conducted detailed interviews in three target application areas in order to (a) assess current practices in community engagement and behavioral intervention, and (b) evaluate the need for and potential benefits of technology. For a fourth area, we did an online survey of coordination with volunteers in order to extend the possible applications into more complex outcomes.

### 4.1.1 Method

Personnel at BranchTime conducted face-to-face, open-ended interviews with practitioners in three target application areas: (i) academic support for disadvantaged high school students, (ii) vocational training for adults, and (iii) outreach to parents of preschoolers in non-English speaking homes. A fourth area -- volunteer coordination -- was investigated by surveying existing websites/apps and talking to volunteer coordinators in a few domains. These areas were selected to address three criteria:
- representativeness of the general category of active/social-support networks;
- distinctness of the particular populations served; and
- potential breadth of social impact if better support methodologies can be established.

Particular interviewees were selected at institutions in the Bay Area for convenient access, and recruited through the authors' informal professional network.

Interviewees were informed that the goals were to understand their current practices, and to identify problems and shortcomings in their existing communication methods, which could be addressed in follow-up studies integrating digital technologies into their practices.

**4.1.2 Results**

Below we report the main qualitative findings from the three target-domain interviews, and from the informal survey of volunteer coordination. Participants' and their organizations' identities have been anonymized for this paper. Interviews that took place over more than one session with a given organization are referred to as one interview in the summaries below.

1. *Academic support for high school students*. The first interview was with a teacher who, in addition to her other teaching duties, does academic coaching in a San Francisco Bay area high school. This school operates a program supporting promising but underperforming students, mostly from disadvantaged or minority backgrounds, who would like to enter college. In most cases they would be the first in their families to do so. Prior to our interview, the program made heavy use of text messaging to provide personalized support to each student: "You need to talk to your chemistry teacher to get your grade up," or "I need to see a draft of your personal statement - your application is almost due." The main problems identified were that the text-message approach (a) placed a heavy burden on the coach and (b) provided limited ability to add stakeholders, collect feedback about progress, and automate follow-up nudges. Improvements in these areas were identified as goals for technology intervention.

2. *Vocational training for adults*. The second interview was with staff at a Bay area vocational training site. Here the emphasis is on augmenting life-skills counseling via real-world nudges, timed for maximum impact on program graduates seeking jobs, showing up for interviews, or joining a new project or work environment. The main problems identified were that prior attempts to communicate with trainees, using conventional email, were marked by low effectiveness and poor response rate. The goals identified for technology application were to improve on these outcomes using mechanisms of recurrent nudges, better feedback loops, and accountability.

3. *Outreach to parents of preschoolers*. The third interview was in the area of early learning, at a Bay area county organization attempting to improve outreach to parents of pre-kindergarten students in non-English-speaking homes. There is well-established research on the effectiveness of text messaging to encourage parents of preschoolers to read to their children and play educational games (York & Loeb, 2014). However, as in the case of high school coaching, this interview revealed that the effectiveness of these interventions is limited by issues of personalization, recurrent engagement of attention, and contexts of accountability. The identified goal for technology was to demonstrate behavioral changes that have been difficult to achieve with prior methods.

4. *Volunteer coordination for citizen science*. The fourth area of interest for application was volunteer coordination. Many organizations have an ongoing interest in attracting volunteers among community members to lend their ideas and energy in service of the organization's mission. Websites have been developed to facilitate this process.[6] Our investigations indicate, however, that these work best when tasks are defined in advance centrally and then "claimed" by would-be volunteers, under the loose supervision of the coordinator. Many would-be volunteers are reluctant to commit time to tasks that are generic in nature and make little use of their special skills and interests. Social prompting infrastructure (see section 5), on the other hand, holds out the promise of organically emerging clusters of activity, well matched to the skills and interests of the volunteers, where the coordinator plays the role of recruiter, matchmaker, and facilitator. This kind of organization would not be possible if the activation energy required for each step of each subprocess were high. Our initial investigation of this style of volunteer coordination identified citizen science, with clusters of serious amateur scientists self-organizing around mini-projects of interest, as a target application area for coordination technology. One goal is to address scalability problems associated with teams (see Staats, Milkman, & Fox, 2012).

---

[6] Examples include VolunteerMatch (http://www.volunteermatch.org/) and linkAges TimeBank (https://timebank.linkages.org/).

## 4.2 Survey of Relevant Software

Our investigation of previously/currently available software focused on three types of tools relevant to social service providers and coordination: (1) tools aimed at providing information for professional network members, (2) ones aimed at sharing information and guidance with community members, and (3) ones aimed at supporting social project or task management.

### 4.2.1 Tools for Sharing Information in Networks

As was mentioned in section 3, networks of individuals sometimes exist primarily for information and knowledge sharing. Examples include professional networks or communities of practice (Wenger, 1998).

There are many tools that support coordination for such information-sharing, passive support networks. For example, in the academic realm, sites such as Google Scholar, Academia.EDU, SSRN, and ResearchGate allow researchers to share papers, see who is interested in similar topics, and learn about projects, funding, conferences, journals, associations, and other resources of mutual interest.[7] Professional networking sites such as LinkedIn[8] allow people in various fields to see who works in particular industries or for particular companies or organizations, and to share other information as well. But these sites do not generally support coordination across networks for concrete, time-dependent actions that have interlocking dependencies, of the type required in active support networks. ResearchGate is taking steps in this direction by allowing researchers to define projects and to request feedback from others, but this is a specialized application not designed for social service networks. Hence, we find these tools not to be particularly useful for the types of ASNs we are considering.

### 4.2.2 Tools for Sharing Information With Targeted Communities

In recent years, a host of digital tools have emerged that attempt to provide more ongoing social support and behavioral guidance to community members in specific areas. For example, teachers use specialized classroom management tools that draw on conventions of social media to organize in-class projects, and then include information in a parent-facing portal so that family members can be informed and find ways of augmenting the teacher's activities (York & Loeb, 2014). Similarly, in healthcare, numerous portals have appeared that give patients access to information relevant to their condition, and to a medium for family and friends to join an extended care circle.[9]

All of these systems are beneficial as far as they go, but they suffer from architectural limitations that will require significant redesign if deeper impact is to be achieved.

First, the behavioral intervention mechanism operates as an open loop, or at a cycle time not well matched to the behavior being targeted. An example is the physician who instructs her patient at the end of an appointment to take certain medications, see a specialist, lose ten pounds, and see her again in four months. Research shows that up to 50% of such instructions are forgotten within sixty minutes of the appointment. Compliance is verified at the next appointment (Selic *et al.*, 2011).

A second problem has to do with the number of websites and apps that offer support for a community member in a specialized area of life. These systems compete for attention both within a category and across categories, require that would-be users -- and members of potential active support networks -- learn specialized workflows, and generate large numbers of notifications and triggers that are not coordinated with each other or with the user's life rhythms. This is sometimes referred to as the "interoperability problem" (Blair & Grace, 2012).

---

[7] See http://scholar.google.com, http://www.academia.edu, http://www.ssrn.com, and http://www.researchgate.com.
[8] See http://www.linkedin.com.
[9] For example, Kaiser Permanente Thrive (https://thrive.kaiserpermanente.org/).

A third problem has to do with the static nature of these apps and websites. Because they are developed by software designers and programmers, and because they do not provide a general-purpose model of the user's emerging and nested goals and the dynamic social support relations that accompany them, they are brittle, which limits their effectiveness in guiding decision and action.[10]

### 4.2.3 Tools for Social Project or Task Management

There is a category of software - "task management" or "project management" - that allows individuals or members of collaborative teams to create tasks, keep track of their completion, and, in team-based tools, assign them to teammates, exchange comments, and so on. Examples we found in current use include Wunderlist (Pro), Asana, Trello, Basecamp, and SAP Jam,[11] but a large number of such systems have been developed since the 1980s, e.g. The Coordinator, which was aimed at project teams in offices (Flores *et al.*, 1988) as well as tools developed within the Center for Coordination Science at MIT (Malone & Crowston, 1994).[12]

These tools are primarily intended for use in the workplace, though some (like Trello) have a significant following in educational settings, and Wunderlist has a free personal version. As a result, there is significant overlap (though not complete identity), depending on the specific system, between the models of task ownership and participation, task hierarchy, and data (due date, completion status, etc.) in these tools. The main differences we found between these tools and what is needed in social-service support networks arise from the differing requirements of work situations and everyday life. In work situations, there is often a need for administrative control, hierarchical management and inspectability. Further, the work context creates a dynamic of task engagement that is governed by the norms, culture, authority relations, incentives, and sanctions of the workplace.

These attributes of the workplace differ considerably from the dynamics of everyday life, where there is less homogeneity, more casual participation, fewer guarantees of timely engagement, and where personal and social incentives must carry most of the burden for performance. As one measure of this, in each of the above cited task management tools that allow a task to be given (assigned) to someone else, there appears to be no requirement that the assignee explicitly accept the task in order for it to be added to their to-do list. In addition, task sharing in these tools is generally tied to a team list that is defined independently of the task, instead of being easily and flexibly controllable on a task-by-task basis. Both of these features are consistent with teams that are defined by larger projects or organizations, and members who have an obligation to their teams. But in active support networks, as we have defined them, each action of support by an individual in the network is voluntary, and those who request help or who suggest actions often have no way to enforce their expectation. Tool design must reflect these realities if it is to work effectively for the looser social structure of an ASN.

### 4.2.4 Software Survey Conclusions

From this survey of existing software, we have concluded that what is needed for social-service support networks is a tool that combines elements of project management with the flexibility to engage based on a user's availability, interest, and convenience. We have developed a model (see section 5) that incorporates *social prompting* and other elements designed to engage users at moments of need and availability. The social prompting model addresses the use cases of social-service support by establishing a master habit of engagement, into which can be slotted particular goals and processes, and in providing explicit triggers for cognitive operations involved in planning, delegating, committing to time frames, and following through.

---

[10] Brittleness has long been recognized as a problem in software and AI (see e.g. Lenat, Prakash, & Shepherd, 1985).
[11] See https://www.wunderlist.com/pro/, https://asana.com/, https://trello.com/, https://basecamp.com/, and http://go.sap.com/product/content-collaboration/enterprise-social-collaboration.html.
[12] See http://ccs.mit.edu/ccsmain.html.

A further conclusion is that if software aimed at individuals outside of defined organizations or teams made it easy to suggest and share tasks with others, then people who are not in a professional support role would be better able to share their knowledge and skills with those who can use them.

## 5. A Framework for Social Prompting and Coordination

We are interested in systems that explicitly store and manipulate goal-related information that individuals (a) carry and process mentally and (b) use in order to move more effectively, through episodes of engagement, from intention to completion. The simplest such systems are ordinary to-do lists, task managers, and calendars. These systems record decisions about events and actions, but they do not themselves take an active role in eliciting the fleeting subordinate events along the way.

From our perspective, the larger opportunity is to create a technology with the following attributes:
- It is general-purpose, allowing users to apply it to goals across different domains of interest and levels of abstraction.
- It is scalable, within the operating limits of human cognition and communication.
- It is psychologically realistic, in that it is specifically designed to conform to attributes of attention, recall, incremental unpacking of plans, and so on.

In this section we describe an open architecture to address these requirements. Illustrated in Figure 1, it consists of three primary elements.

The first element is a *coordination engine*, which manages a data structure known as the *activity graph*. The activity graph consists of a collection of objects, known as *tasks* (which could alternatively have been called goals, activities, commitments, intentions, etc.).

The tasks are organized by two kinds of relations: *social* and *logical*. Social relations connect each task to its *owner* (responsible party) and to any other *participants* who are part of the owner's active support network for that task. Logical relations represent instrumental relationships between goals and subgoals.

The coordination engine interprets coordination protocols that, among other things, allow for creation of new tasks, invitation of new participants, hand-off of tasks (via two-step offer and acceptance), marking of completion or abandonment of tasks, spawning of subtasks, and so on. The engine also contains algorithms that analyze patterns of engagement of individuals with tasks, and attempts to optimize the frequency and timing of engagement by users with active tasks.

The presentation of tasks is mediated by a software utility called the *personal prompter*, the second component of the architecture. The user sets aside short sessions, a few minutes in length, several times a day, usually in response to recommendations by the coordination engine. During those sessions, the prompter displays items that require the user's attention, and provide affordances for quickly registering decisions and short communications to other participants involved in each relevant item.

The idea of the prompter is to rationalize the flow of attention, decision, and action, and to provide points of synchronization and control. Ordinarily, a user would not carry out a named action during the prompter session, unless it was very short, or involved an operation intrinsic to the prompter itself. Rather, the user would be primed during the prompter session to take a number of actions in the interval between the end of the prompter session and the next session. The timing of sessions would be personalized to maximize the benefit to the user. This differs from asynchronous message systems such as email, as well as workflow systems such as those of Asana.com and The Coordinator (Flores *et al.*, 1988), which inform a user about tasks to be performed, but do not schedule the user's engagement purposefully at pre-selected optimal moments.

The final component of the architecture is the *template library*, which stores reusable plan fragments in parametric form, thereby providing a mechanism for sharing executable, procedural knowledge among users. The simplest templates would be collections of action steps, but with owners and other task participants represented by role variables, rather than named individuals. To use the template, a user would load it into their personal prompter, optionally modify steps or roles, associate roles with actual named users, and click "Launch." The system would push instantiated tasks to participants, inserting them into the prompter cycle of those individuals. The intention is for templates in the template library to be authored and curated by members of the community as an open-source resource, covering a wide range of situations in everyday life., e.g., planning a trip to Paris, preparing a will, helping a loved one transition to assisted living, getting into college, etc. These scenarios are characterized by informal collaboration involving family members, friends, and professionals.

**Fig. 1** An architecture for social prompting and coordination

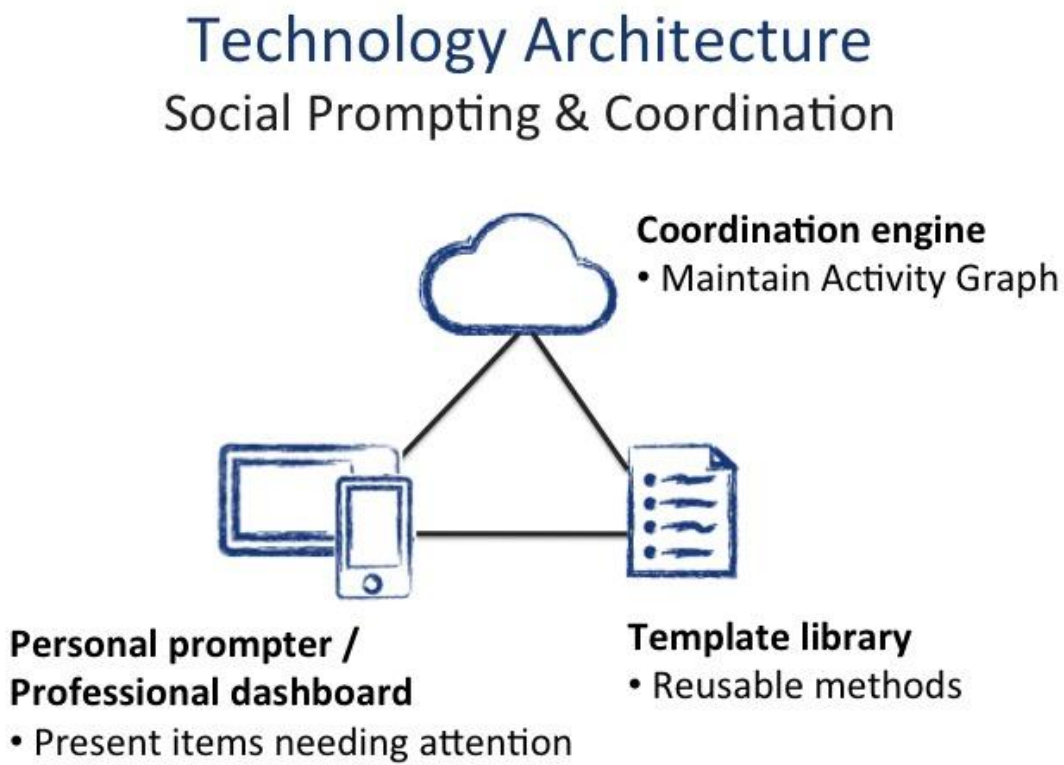

**6. The Design in Action: An Illustrative Example**

The architecture described above has been implemented in a tool that is available at http://www.branchtime.com/. We can illustrate how it operates in a social-service support network through a simple example.

Imagine that a job placement counselor named Stan is working with his client, Alex, who is trying to find a job. Stan suggests that Alex access BranchTime on his smartphone, and Alex does so. When Alex creates an account and begins to use the software, he is prompted to create a schedule of daily times when he will open his prompter and engage with a set of goals (see Fig. 2). In the figure, Alex schedules sessions at 6pm, 1pm, and 11 am. These sessions represent the times when Alex will be prompted to think about, plan, and act on his goals. He is alerted with a text message at the times he chooses. The self-chosen times reflect that these are voluntary times of engagement for Alex, when he anticipates having available time, but through notification, the technology helps Alex to stay on track to follow-through at her chosen times -- overcoming a problem Alex may be more likely to have outside of an organized work setting, and which was identified in section 4, namely that there is no formal accountability mechanism requiring Alex to follow through on his goals.

**Fig. 2** A phone screen showing the scheduling of engagement sessions

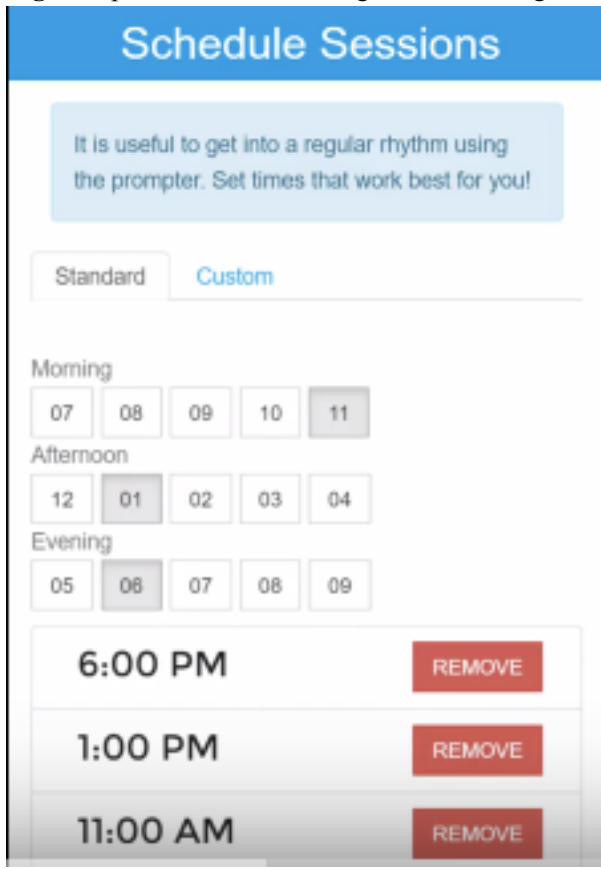

Stan has urged Alex to create a presentation about his past work, which he can show to prospective employers. With Alex now set up on the software, Stan is able to create a goal – "Create presentation" – and to hand it off to Alex. Alex sees the goal on his phone as he reviews handoffs, and is given the option to Accept or Decline the goal (Fig. 3). If Alex accepts the goal, the software prompts him further to "Break it down" into subgoals or actionable tasks. Stan can also hand off more complex templates (which are more like projects) to Alex, which not only specify a goal but a sequence of steps for completing it.

Once Alex has accepted the goal from Stan, Alex is the Owner of the goal. And while reviewing the goal, Alex can add further participants (Fig. 4). Imagine that Alex has a friend, Jennifer, who is an expert on presentations. Alex wants to ask Jennifer to review his presentation after Alex creates it. Alex adds Jennifer as a participant on the "Create presentation" goal, which will allow Jennifer to keep track of his progress on each of the sub-goals or tasks into which Alex breaks down the goal of creating a presentation. When the presentation is ready for Jennifer's review, her prompter will let her know that she can review what Alex has created, and she can send her feedback to Alex either with a comment attached to the task, or through other communication (e.g. by email). This illustrates how a volunteer can be added, in this case by the client of a professional job counselor, and becomes part of the client's active support network even though there is no formal or organizational relationship between Stan and Jennifer.

The software helps Jennifer to help her friend Alex, by making it easier for her to see exactly when her help is needed and by adding the review task as a goal if she merely accepts Alex's invitation. She does not have to remember to do this on her own, which might make it more likely that Jennifer, especially if she is a busy professional, would fail to do the task. This illustrates how BranchTime attempts to facilitate contributions to the social good for which there is much under-utilized capacity.

**Fig. 3** Reviewing a proposed goal handoff from another user

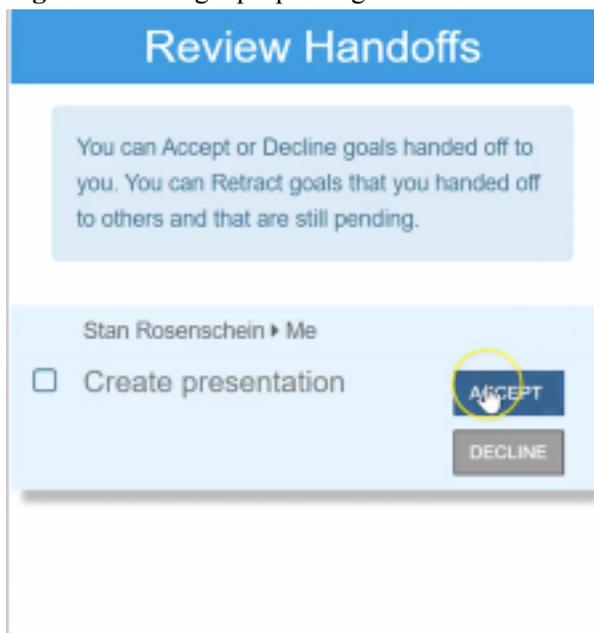

The software contains a number of other features, e.g. for commenting, and for classifying the goal according to whether it can be done immediately or requires further refinement. But for understanding how it is used in social-service support, the key elements of the design are the following:
1. *Multiple stakeholders*. A task (or goal) can have more than one participant in addition to the owner, who will be completing it. This facilitates both communication and support in the interpersonal network.
2. *Handoffs*. In contrast to an "assignment", a handoff must be explicitly accepted or declined by the prospective owner of a task. This allows the professional job counselor, in the above example, to place items in the goal list of a client, but in a way that respects the client's autonomy to decide whether and how to act.
3. *Scheduled prompting sessions*. Choosing times to engage with the prompter, with reminders at those chosen times, helps the client in this example to stay on track outside of an otherwise structured work schedule, despite the freedom the client has to skip the engagement. The scheduled times also facilitate coordination through communication with other participants, who can anticipate when tasks will be completed and schedule their own engagement accordingly.
4. *Flexibly controllable social updating*. Both the counselor and the client are easily able to specify who is a participant on each task, as well as which updates (e.g. task creation, handoffs and responses, task status changes, and completions) are shared and with whom. This both respects the individual autonomy of each user in an active support network, and facilitates the fluid relationships that exist in such networks, in which new participants come and go and relationships are often configured on a task-by-task basis.[13]

**Fig. 4** Adding participants to a goal

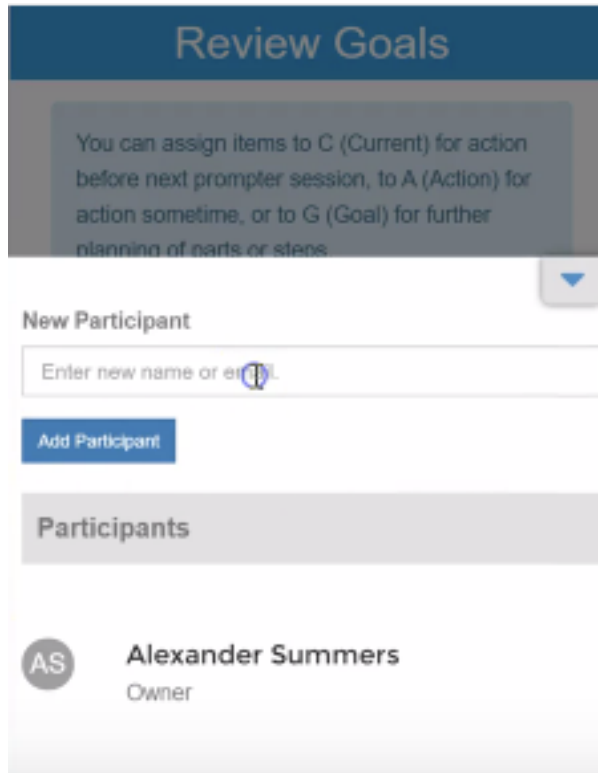

The above example shows how the features of the design in section 5 respond to active support network users' software needs identified in section 4. Perhaps the most important point is that coordination in networks, as opposed to teams, as defined in section 3, requires features that respect the autonomy of individual network members and the fluid, fine-grained nature of the relationships between members. Giving users more control over the acceptance of tasks given to them by others, as well, as more precise control over what information is shared, also addresses problems of task and information overload that come from unsolicited and/or excessive task assignment and information sharing.

**7. Conclusions and Future Work**

The interviews and need-finding investigations reported here have confirmed our belief that there is enough good will and raw capacity to produce a great deal of social good. In each of our interviews, we identified coordination that could be facilitated by greater social prompting as the weak link. Traditional communication methods are leaky. They squander a large fraction of available good will and capacity.

---

[13] Social updating may also make goal accomplishment more likely under the public commitment effect when the goal owner desires the outcome (Staats, Harland, & Wilke, 2004; Hollenbeck, Williams, & Klein, 1989), but there is also a danger that announcing one's intentions will reduce motivation to achieve them if the goal is viewed more as a signal of identity than as truly desired (Gollwitzer *et al*., 2009).

Specific technologies, psychologically realistic and optimized for social coordination, might propel social progress dramatically.

The architectural design described in section 5 has been implemented in a working system that is currently being pilot-tested with existing networks of support. Our main learnings from discussions with prospective users during this phase have been the following:
(1) The importance of social buy-in by professionals for initial adoption in the scenarios addressed by coordination technology.
(2) The importance of templates to provide packaged solutions to common situations to insure sufficiently short path to payoff and reinforce adoption.
(3) General receptivity to the prompting solution, and acknowledgement that persistent follow up and coordination among stakeholders is a real problem in everyday life and in the scenarios investigated.

We are currently beginning follow-up studies of the use of this technology at the organizations interviewed as part of our needfinding investigation. These studies will help us to understand more details about the challenges of coordination technology applied to the target areas, as well as the commonalities that characterize these areas of social good for designers of coordination tools.

**References**

Arrow K (1963) Social choice and individual values, 2$^{nd}$ edition. Yale University Press, New Haven

Baron J (2007) Thinking and deciding, 4th edition. Cambridge University Press, Cambridge, UK

Bazerman M, Neale MA (1992) Negotiating rationally. The Free Press, New York

Davies T, Gangadharan SP, eds (2009) Online deliberation: Design, research, and practice. CSLI Publications, Stanford

De Cindio F, Klein M, De Liddo A, Schuler D, Buckingham Shum S, original signatories (2014) Collective intelligence for the common good community / network statement of principles. Public Sphere Project. http://www.publicsphereproject.org/content/statement-principles-collective-intelligence-common-good-community-network. Accessed 5 November 2016

Flores F, Graves M, Hartfield B, Winograd T (1988) Computer systems and the design of organizational interaction. ACM Trans Office Info Systems 6:153-172. doi:10.1145/45941.45943

Fogg BJ, Hreha J (2010) Behavior wizard: a method for matching target behaviors with solutions. Persuasive technology 117-131. Springer, Berlin Heidelberg

Gawande A (2010) The checklist manifesto: How to get things right. Metropolitan Books, New York

Godin G, Conner M, Sheeran P (2005) Bridging the intention–behaviour gap: The role of moral norm. British J Soc Psych 44:497-512. doi:10.1348/014466604X17452

Gollwitzer PM, Sheeran P, Michalski V, Seifert AE (2009) When intentions go public: Does social reality widen the intention-behavior gap? Psych Sci 20:612-618. dos:10.1111/j.1467-9280.2009.02336.x

Grace P, Blair G (2012) Emergent middleware: Tackling the interoperability problem. IEEE Internet Computing 16:78-82. doi:10.1109/MIC.2012.7

Greenfield A (2013) Against the smart city. Amazon, Seattle

Habermas J (1984) Theory of communicative action, Volume one: Reason and the rationalization of society. Tr. by McCarthy TA. Beacon Press, Boston, MA

Habermas J (1987) Theory of communicative action, Volume two: Lifeworld and system: A critique of functionalist reason. Tr. by McCarthy TA. Beacon Press, Boston, MA

Hollenbeck JR, Williams CR, Klein HJ (1989) An empirical examination of the antecedents of commitment to difficult goals. J Appl Psychology 74:18-23. doi:10.1037/0021-9010.74.1.18

Kuo FY, Young ML (2008) A study of the intention–action gap in knowledge sharing practices. J Amer Society Info Sci and Tech 59:1224-1237. doi:10.1002/asi.20816

Lenat DB, Prakash M, Shepherd M (1985) CYC: Using common sense knowledge to overcome brittleness and knowledge acquisition bottlenecks. AI Magazine 6:65. http://www.aaai.org/ojs/index.php/aimagazine/article/viewArticle/510. Accessed 9 November 2016

Levitin DJ (2014) The organized mind: Thinking straight in the age of information overload. Penguin, London

Malone TW, Crowston K (1994) The interdisciplinary study of coordination. ACM Computing Surveys 26:87-119. doi:10.1145/174666.174668

Olson M (1971) The logic of collective action: Public goods and the theory of groups, Revised edition. Harvard University Press, Cambridge, MA

Rogers T, Milkman KL, Volpp KG (2014) Commitment devices: using initiatives to change behavior. JAMA 311:2065-2066. doi:10.1001/jama.2014.3485

Sadowski J (2016) Selling smartness: Visions and politics of the smart city. Dissertation, Arizona State University. https://repository.asu.edu/attachments/175031/content/Sadowski_asu_0010E_16271.pdf. Accessed 8 November 2016.

Searle J (1969) Speech acts. Cambridge University Press, Cambridge, UK

Selic P, Svab I, Repolusk M, Gucek NK (2011) What factors affect patients' recall of general practitioners' advice?. BMC Family Practice 12:141. doi:10.1186/1471-2296-12-141

Sniehotta FF, Scholz U, Schwarzer R (2005) Bridging the intention–behaviour gap: Planning, self-efficacy, and action control in the adoption and maintenance of physical exercise. Psychology & Health 20:143-160. doi:10.1080/08870440512331317670

Staats H, Harland P, Wilke HA (2004) Effecting durable change: A team approach to improve environmental behavior in the household. Environ & Beh 36:341–367. doi:10.1177/0013916503260163

Staats BR, Milkman KL, Fox CR (2012) The team scaling fallacy: Underestimating the declining efficiency of larger teams. Org Behavior & Human Decision Processes 118:132-142. doi:10.1016/j.obhdp.2012.03.002

Sunstein C, Thaler R (2008) Nudge: Improving decisions about health, wealth, and happiness. Yale University Press, New Haven

Tetlock P, Mitchell G (2009) Implicit bias and accountability systems: What must organizations do to prevent discrimination? Res Org Behavior 29:3-38. doi:10.1016/j.riob.2009.10.002

Vermeir I, Verbeke W (2006) Sustainable food consumption: Exploring the consumer "attitude–behavioral intention" gap. J Ag and Environ Ethics 19:169-194. doi:10.1007/s10806-005-5485-3

Wenger E (1998) Communities of practice: Learning, meaning, and identity. Cambridge University Press, Cambridge, UK

York BN, Loeb S (2014) One step at a time: the effects of an early literacy text messaging program for parents of preschoolers. doi:10.3386/w20659